\documentclass[a4paper, conference]{IEEEtran}

\usepackage{amsmath, bm}

\usepackage{graphicx,color}

\newcommand{\conforfull}[1]{See Appendix #1 for the proof.}

\newtheorem{theorem}{Theorem}
\newtheorem{lemma}{Lemma}
\newtheorem{corollary}{Corollary}
\newtheorem{proposition}{Proposition}
\newtheorem{pr}{Property}

\newcommand{\defeq}{ := }

\newcommand{\beqa}{\begin{eqnarray}}
\newcommand{\eeqa}{\end{eqnarray}}
\newcommand{\beqno}{\begin{eqnarray*}}
\newcommand{\eeqno}{\end{eqnarray*}}

\begin{document}
\title{ 
%
%
An Iterative Algorithm for Computing the Optimal Exponent of Correct Decoding Probability for Rates below the Rate~Distortion Function 
} 

\author{
\IEEEauthorblockN{Yutaka Jitsumatsu}
\IEEEauthorblockA{
Dept. of Informatics, \\
Kyushu University, Japan\\
Email: jitumatu@inf.kyushu-u.ac.jp}
\and
\IEEEauthorblockN{Yasutada Oohama}
\IEEEauthorblockA{
Dept. of Communication Engineering and Informatics, \\
University of Electro-Communication, Japan\\
Email: oohama@uec.ac.jp}
}

\maketitle

\begin{abstract}
The form of 
Dueck and K\"orner's exponent function for 
correct decoding probability for discrete memoryless channels 
at rates above the capacity 
is similar to 
the form of 
Csisz\'ar and K\"orner's exponent function for 
correct decoding probability in lossy source coding 
for discrete memoryless sources
at rates below the rate distortion function. 
We recently gave a new algorithm for computing
Dueck and K\"orner's exponent. 
In this paper, we give an algorithm for computing 
Csisz\'ar and K\"orner's exponent. 
The proposed algorithm can also be used to compute
cutoff rate and the rate distortion function.


\end{abstract}


\IEEEpeerreviewmaketitle

\section{Introduction}
Computation of the channel capacity 
of a discrete memoryless channel (DMC) under input constraint 
and computation of the rate distortion function of a 
discrete memoryless source (DMS) have similar 
structure{s}~\cite{Blahut1972}\cite[Chapter 8]{Csiszar-KornerBook}.
Algorithms for computing channel capacity were given by 
Blahut~\cite{Blahut1972} and Arimoto~\cite{Arimoto1972} 
and an algorithm for computing rate distortion function
was given by Blahut~\cite{Blahut1972}.

For channel coding, a strong converse theorem was established 
by Wolfowitz~\cite{Wolfowitz}.
Arimoto proved that the probability of correct decoding 
vanishes exponentially if the transmission rate 
is above the capacity~\cite{Arimoto1973}.
{He then gave an algorithm for computing his exponent 
function~\cite{Arimoto1976}.
Subsequently,} Dueck and K\"orner gave the optimal exponent function 
of {correct decoding probability}~\cite{Dueck_Korner1979}. 
They claimed that their exponent function {coincides with}
Arimoto's exponent. However, 
the forms of these two exponent functions are quite different.
We recently proposed an algorithm for computing 
Dueck and K\"orner's exponent function~\cite{OohamaJitsumatsuISIT2015}. 
{
The difference between Arimoto's algorithm and the
recently proposed one is as follows:
In Arimoto's algorithm, 
the probability distribution over the input alphabet 
and backward transition distribution are updated alternately.
On the other hand, 
in the proposed method, 
joint probability distribution over input and
output alphabets is iteratively updated. 
}

For source coding for DMSs, 
the rate distribution function, denoted by $R(\Delta|P)$, 
indicates the minimum admissible rate at distortion level 
$\Delta$ for a source with distribution $P$.
{The source coding 
theorem under $\epsilon$-fidelity criterion 
states that 
if the coding rate $R$ is above $R(\Delta|P)$, 
the probability of an event in which 
the distortion measure between input sequence
and its reproduced one 
exceeds $\Delta$ tends to zero exponentially~\cite{Blahut1974, Marton1974}. 
In~\cite{Arimoto1976}, 
an algorithm for computing 
an exponent function of this probability has also been given. 
On the other hand, 
the strong converse theorem states that
the probability of an event in which 
the distortion measure exceeds $\Delta$ tends to one 
if $R<R(\Delta|P)$. 
The optimal exponent for $R<R(\Delta|P)$ was determined by
Csisz\'ar and K\"orner~\cite{Csiszar-KornerBook}. 
This exponent function is expressed by a form similar to
the form of 
Dueck and K\"orner's exponent function {for channel coding}.
An algorithm for computing the exponent of correct
decoding probability for the rates $R<R(\Delta|P)$ has 
not been provided. 
}

In this paper, we give an {iterative} 
algorithm for computing 
Csisz\'ar and K\"orner's exponent function. 
The algorithm 
has a structure similar to our recently proposed
algorithm for computing 
{Dueck and K\"orner's exponent 
function}~\cite{OohamaJitsumatsuISIT2015}. 
We give a proof {in which the probability distribution
computed by the algorithm converges to the optimal distribution. }
We also show that the proposed algorithm can be used to compute
{cutoff rate and}
the rate distortion function. 

{
Developing a new algorithm for 
computing the correct decoding probability exponent 
in lossy source coding has a limitted practical importance because 
the strong converse theorem already states that 
correct decoding probability goes to zero if the coding rate 
is below the rate distortion function. 
The correct decoding exponent expresses how fast
such a probability goes to zero.
However, analyzing the correct decoding probability exponent 
and comparing it with the error exponent in source coding as well as 
the one in channel coding brings 
a better understanding of the structure of these exponent functions.
In addition, the results of this paper may 
lead to the development of a computation algorithm for 
other coding schemes.
}

\section{Source coding at rates below the rate-distortion function}

{
This section gives definitions for quantities that are necessary
to describe the correct decoding probability exponent of
sournce coding for discrete memoryless sources (DMSs)
at rates below the rate distortion functions. 

Let $\mathcal{X}$ be 
a source alphabet and $\mathcal{Y}$ be 
a reproduction alphabet. Both $\mathcal{X}$
and $\mathcal{Y}$ are supposed to be finite.
A $k$-length block code for sources with alphabet
$\mathcal{X}$ is a pair of mappings 
$(\varphi^{(k)}, \psi^{(k)})$, where  
$\varphi^{(k)}$ is an encoding function
that maps every element of $\mathcal{X}^k$
into $\mathcal{M}_k = \{1, 2\ldots, |\mathcal{M}_k|\}$ in a one-to-one manner
and 
$\psi^{(k)}$ is a decoding function
that maps every element of $\mathcal{M}_k$ 
into $\mathcal{Y}^k$,
where 
$\mathcal{M}_k$
is an index set. 
The rate of such a code is defined as 
$\frac1k \log | \mathcal{M}_k|$. 
Let $d(x,y)\geq 0$ be a distortion measure
for $x\in \mathcal{X}$ and $y \in \mathcal{Y}$.
The average distortion between $x^k$ and $y^k$
is defined as
$d(x^k, y^k) = \frac1k \sum_{i=1}^k d(x_i, y_i)$.
We assume that for every $x \in \mathcal{X}$, 
there exists at least one $y \in \mathcal{Y}$ such
that $d(x,y) = 0$.

Let $P$ be a probability distribution over source alphabet $\mathcal{X}$. 
%
%
Correct decoding is defined as an event {in which} the 
distortion does not exceed a prescribed distortion level $\Delta$. 
We denote the probability of correct decoding 
by $P_{\rm c}^{(k)}(\varphi^{(k)}, \psi^{(k)}; \Delta| P)$, 
which is given by
\begin{align*}
P_{\rm c}^{(k)}(\varphi^{(k)}, \psi^{(k)}; \Delta| P)
=
\mathrm{Pr}\{
d( X^k, \psi^{(k)} ( \varphi^{(k)} (X^k)) )
\leq \Delta
\}. 
\end{align*}
}
\newcommand{\AchievableRare}{
The average probability of an event that
the distortion exceeds $\Delta$ 
is given by $P_{\rm e}^{(k)}(\varphi^{(k)}, \psi^{(k)}; \Delta| P)
= 1 - P_{\rm c}^{(k)}(\varphi^{(k)}, \psi^{(k)}; \Delta| P)$.

A rate $R$ is said to be $\varepsilon$-achievable
at distortion level $\Delta$ if 
for every $\delta>0$,
and sufficiently large $k$ there exist $k$-length
block codes $(\varphi^{(k)}, \psi^{(k)})$
satisfying
\begin{align*}
&\frac1{k} \log || \varphi^{(k)} || \leq R + \delta, \\
&{\rm Pr} \{ d(X^k, \varphi^{(k)} ( \psi^{(k)} ( X^k ) ) )
\leq \Delta \}
\geq
1-\varepsilon,
\end{align*}
{where $|| \varphi^{(k)} ||$ denotes 
the cardinarity of the range of $ \varphi^{(k)} $. 
}
Define 
\begin{align*} 
R_{\varepsilon}(\Delta|P) &\defeq \inf 
\{ 
R: R \text{ is $\varepsilon$-achievable}  \}, \\
R(\Delta|P) &\defeq \inf
\{ R: R \text{ is $\varepsilon$-achievable
for $\forall \varepsilon \in (0,1)$}\}.
\end{align*} 



The following theorem is stated in~\cite[p.109]{Csiszar-KornerBook}: 

\begin{theorem}
For $\forall \varepsilon \in (0,1)$ and $\Delta \geq 0$, 
we have 
$$
R_{\varepsilon}(\Delta | P) = 
R(\Delta| P) = \min_{
\scriptstyle 
V \in \mathcal{P}(\mathcal{Y}|\mathcal{X}): 
\atop 
\scriptstyle 
{{\rm E}_{(P,V)}[d(X,Y)]}
\leq \Delta
} 
I(P, V),$$ 
where
$\mathcal{P}(\mathcal{Y}|\mathcal{X})$ is
a set of conditional {probability distributions on $\mathcal{Y}$ given 
$\mathcal{X}$}. 
\end{theorem}

}

{
The exponent of the maximum of 
$ P_{\rm c}^{(k)}(\varphi^{(k)}, \psi^{(k)}; \Delta|P) $ 
over all pairs of encoding and decoding 
functions {having a rate} less than $R$
is defined by}
\begin{align*}
& G^{(k)} (R, \Delta | P) \\
& \defeq 
\min_{
\scriptstyle 
 (\varphi^{(k)}, \psi^{(k)}): 
\atop
\scriptstyle 
\frac1k \log |\mathcal{M}_k| \leq R 
}
{\textstyle 
\left(
- \frac1k
\right)
}
\log 
P_c^{(k)}(\varphi^{(k)}, \psi^{(k)}; \Delta| P).
\end{align*}
Let 
\begin{align*}
&G^* (R, \Delta|P) = \lim_{k\to \infty}
G^{(k)}(R, \Delta|P).
\end{align*}
The optimal exponent 
$G^*(R, \Delta|P)$ is 
determined by Csisz\'ar and K\"orner 
in~\cite[p.139]{Csiszar-KornerBook}.
%
\newcommand{\PrOptimalG}{
This function satisfies the following property.

\begin{pr}\label{pr:optimalG}
\

\begin{itemize}
\item[a)] For a fixed $\Delta \geq 0$, $G^{(k)}(R, \Delta|P)$
is a monotone decreasing function of $R \geq 0$.
For a fixed $R\geq 0$, $G^{(k)}(R, \Delta|P)$ is a 
monotone decreasing function of $\Delta \geq 0$. 

\item[b)] 
The sequence $\{ G^{(n)}(R, \Delta|P) \}_{n\geq 0}$
of exponent functions satisfies the following
sub-additivity property: 
\begin{align*}
G^{(k+\ell)} (R, \Delta| P)
\leq \frac{
k G^{(k)}(R, \Delta |P) + \ell G^{(\ell)} (R, 
\Delta|P)
}{k+\ell},
\end{align*}
from which we have that $G^{*}(R, \Delta|P)$
exists and is equal to $\inf_{k\geq 0} G^{(k)}(R, \Delta|P)$.

\item[c)]
For a fixed $\Delta\geq 0$, $G^{*}(R, \Delta |P)$ is 
a monotone decreasing function of $R \geq 0$.
For a fixed $R \geq 0$, $G^{*}(R, \Delta |P)$ is 
a monotone decreasing function of $\Delta \geq 0$.

\item[d)] The function $G^{*}(R, \Delta|P)$
is a convex function of $(R, \Delta)$. 

\end{itemize}
\end{pr}

{\it Proof:} 
By definition, part a) is obvious.
By time sharing,
we have that 
\begin{align}
&G^{(k+\ell)} \bigg( \frac{kR + \ell R'}{k+\ell}, 
\frac{k\Delta + \ell \Delta'}{k+\ell} \bigg| P\bigg)\notag\\
&\leq \frac{
k G^{(k)}(R, \Delta |P) + \ell G^{(\ell)} (R', \Delta'|P)
}{k+\ell}
\label{timesharing_CK}
\end{align}
Part b) is the case for $R=R'$ and $\Delta = \Delta'$.
Part c) is obvious from part a) and part b).
Eq.(\ref{timesharing_CK}) implies 
\begin{align}
&G^{*} \left(\alpha R + \bar \alpha R', 
\alpha \Delta + \alpha\Delta'| P\right) \notag\\
&\leq 
\alpha G^{*}(R, \Delta |P) 
+ \bar \alpha G^{*} (R', \Delta'|P).
\end{align}
This proves Part d). \hfill\IEEEQED
}
%
%
In order to describe their result, we 
define 
\begin{align}
&G_{\rm CK}(R, \Delta|P) \notag\\
&\defeq 
\min_{q_X \in {\cal P(X)}  } 
\{   |R(\Delta | q_X) - R|^+ + D(q_X || P) \},
\label{GCK_original}
\end{align}
where {$\mathcal{P(X)}$ is a set of probability
distributions on $\mathcal{X}$, }
$|x|^+ = \max\{0, x\}$, 
\begin{align*}
R(\Delta | q_X) &=
\min_{
\scriptstyle 
q_{Y|X} \in \mathcal{P(Y|X)}: 
\atop
\scriptstyle 
{
{{\rm E}_{q_{XY}}[ d(X,Y)]}
} 
\leq \Delta}
I(q_X, q_{Y|X}),\\
D(q_X || P ) &= 
{\rm E}_{q_X} 
\left[
  \log \frac{q_X(X)}{P(X)}
\right], 
\end{align*}
where $\mathcal{P(Y|X)}$ is a set of conditional 
probability distributions on $\mathcal{Y}$ given 
$\mathcal{X}$. 
Then we have the following theorem: 
\begin{theorem}[Csisz\'ar and K\"orner]
For any $\Delta \geq 0$ and {$0\leq R \leq R(\Delta|P)$, }
we have
\begin{align}
 G^* (R, \Delta | P) = G_{\rm CK} (R, \Delta | P) .
\end{align}
\end{theorem}


The purpose of this paper is to give an
algorithm for computing $ G_{\rm CK}(R, \Delta |P)$.
For this aim, 
we first introduce the following exponent function
and prove it is equivalent to $G_{\rm CK}(R, \Delta |P)$.
We then derive a parametric expression for it. 
Define 
\begin{align}
&  G(R, \Delta|P) \notag \\
& \defeq  
\min_{
\scriptstyle 
q_{XY} \in {\cal P(X \times Y)}: 
\atop
\scriptstyle 
{{\rm E}_{q_{XY}}[ d(X,Y)]}
\leq \Delta 
} \Big \{ 
|I(q_X, q_{Y|X}) - R|^+  + D(q_X || P) \Big \},
\label{GCK}
\end{align}
{where $\mathcal{P(X\times Y)}$ is a set of joint distributions 
on $\mathcal{X} \times \mathcal{Y}$.
}

We have the following lemma:
\begin{lemma}
For any $R, \Delta \geq 0$, we have
$$
 G_{\rm CK}(R, \Delta|P) 
= G(R, \Delta|P). 
$$
\end{lemma}

{\it Proof:} Let $q_{XY}^*$ be
a joint distribution that attains
$G(R, \Delta| P)$.
{From its formula, we have 
\begin{align}
R(\Delta | q_X^*) \leq I(q_X^*, q^*_{Y|X}). 
\label{ineq:lm1}
\end{align}
Thus, }
\begin{align*}
G (R, \Delta|P)
&= |I(q_X^*, q_{Y|X}^*) - R|^+ + D(q_X^* || P)\\
&\stackrel{\rm (a)}{\geq} |R(\Delta | q_X^*) - R|^+ + D(q_X^* || P)\\
& \geq \min_{q_{X} \in \mathcal{P(X)}} \{ |R(\Delta | q_X) - R|^+ + D(q_X || P) \}\\
& = G_{\rm CK} (R, \Delta|P).
\end{align*}
{Step (a) follows from (\ref{ineq:lm1}).}
On the other hand, let $\tilde q_{X}^* $
be a distribution that attains
$G_{\rm CK}(R, \Delta| P)$
and let $\tilde q_{Y|X}^* $ be
a conditional distribution that
attains $R(\Delta|\tilde q_X^*)$. 
Then, we have
\begin{align*}
&G_{\rm CK} (R, \Delta|P) 
= |I(\tilde q_X^*, \tilde q_{Y|X}^*) - R|^+ 
+ D(\tilde q_X^* || P) \\
& \geq \min_{
\scriptstyle 
q_{XY}: 
\atop
\scriptstyle 
{{\rm E}_{q_{XY}}[ d(X,Y)]}
\leq \Delta}
\{ |I(q_X, q_{Y|X}) - R|^+ + D(q_X || P) \} \\
& = G (R, \Delta|P). 
\end{align*}
Thus, we have 
$ G_{\rm CK} (R, \Delta|P) =
G (R, \Delta|P)$, 
which completes the proof.
\hfill\IEEEQED


The function $G(R, \Delta|P)$ satisfies the following 
{property, which is useful for deriving} its parametric expression:
\begin{pr}\label{pr:G_CK}
\
\begin{itemize}
\item[a)] 
$G(R, \Delta|P)$ is a monotone decreasing
function of $R \geq 0$
for a fixed $\Delta \geq 0$ 
and is a monotone decreasing function of $\Delta \geq 0$
for a fixed $R\geq 0$.
\item[b)] $G(R, \Delta|P)$ is a convex function of $(R,\Delta)$.
\item[c)] 
$G(R, \Delta|P)$ takes positive value
for $0 \leq R < R(\Delta|P)$.
For $R\geq R(\Delta|P)$, 
$G(R, \Delta|P)=0$. 
\item[d)]
For $R' \geq R \geq 0$, we have
$
G(R, \Delta | P ) - G( R', \Delta | P )
\leq R' - R.
$
\end{itemize}
\end{pr}


\conforfull{\ref{ProofprII}}

\newcommand{\proofprII}{

\subsection{Proof of Property \ref{pr:G_CK}}
\label{ProofprII}
In this appendix, we prove Property~\ref{pr:G_CK}. 
By definition, Part a) is obvious.
For the proof of Part b), 
let $q^{(0)}$ and $q^{(1)}$ be 
joint distribution functions that attain 
$G(R_0, \Delta_0|P)$ and $G(R_1, \Delta_1|P)$,
respectively. 
Denote 
\begin{align}
&\Theta(R, q | P) 
\defeq |I(q_X, q_{Y|X}) - R|^+ + D(q_X || P) \notag \\
&= \max\{ D(q_X || P),
I(q_X, q_{Y|X}) - R + D(q_X || P) \}.\label{Theta_max}
\end{align}
By definition, we have
\begin{align}
G(R_i, \Delta_i |P) = \Theta(R_i, q^{(i)} |P)
\mbox{ for } i=0,1.
\label{optimal_distribution}
\end{align}
For $\alpha_1 = \alpha \in [0,1]$ and 
$\alpha_0 = 1-\alpha$,
we set 
$R_\alpha = \alpha_0 R_0 + \alpha_1 R_1$, 
$\Delta_\alpha = \alpha_0 \Delta_0 + \alpha_1 \Delta_1$, 
and 
$q^{(\alpha)} = \alpha_0 q^{(0)} + \alpha_1 q^{(1)}$.
By linearity of ${\rm E}_{q}[ d(X,Y)]$ 
with respect to $q$,
we have that
\begin{align}
{{\rm E}_{q^{(\alpha)}} [ d(X, Y) ] }
= \sum_{i=0,1} \alpha_i {{\rm E}_{q^{(i)}} [ d(X, Y) ] }
\leq \Delta_{\alpha}. 
\label{constraint_Delta_alpha}
\end{align}
Because $$I(q_X, q_{Y|X}) + D(q_X||P) = 
\sum_{x,y} q_{XY}(x,y)
\log 
\frac{q_{X|Y}(x|y)}{P(x)}$$
is convex with
respect to $q_{XY}$ and $D(q_X||P)$ is convex with
respect to $q_X$, we have
\begin{align}
& I( q_X^{(\alpha)}, q_{Y|X}^{(\alpha)}) + D(q_X^{(\alpha)} || P) \notag\\
& \leq \sum_{i=0,1} \alpha_i  
\left\{
I( q_X^{(i)}, q_{Y|X}^{(i)}) + D(q_X^{(i)} || P)
\right\},  \label{eq.21a}\\
& D(q_X^{(\alpha)} || P) 
\leq \sum_{i=0,1} \alpha_i 
D(q_X^{(i)} || P).  \label{eq.21b}
\end{align}
Therefore, we have the following two chains
of inequalities:
\begin{align}
& I( q_X^{(\alpha)}, q_{Y|X}^{(\alpha)}) + D(q_X^{(\alpha)} || P) 
  - R_{\alpha} \notag \\
& 
\stackrel{\rm (a)}{\leq} \sum_{i=0,1} \alpha_i  
\left\{
I( q_X^{(i)}, q_{Y|X}^{(i)}) + D(q_X^{(i)} || P)
- R_{i}
\right\} \notag \\
& \stackrel{\rm (b)}{\leq} \sum_{i=0,1} \alpha_i 
\Theta(R_i, q^{(i)} | P), \\
& D(q_X^{(\alpha)} || P) 
\stackrel{\rm (c)}{\leq} 
\sum_{i=0,1} \alpha_i 
D(q_X^{(i)} || P) \notag\\
&\stackrel{\rm (d)}{\leq} 
\sum_{i=0,1} \alpha_i 
\Theta(R_i, q^{(i)} | P).
\end{align}
Steps (a) and (c) follow from (\ref{eq.21a}) and (\ref{eq.21b})
and Steps
(b) and (d) follow from the definition
of $\Theta(R_i, q^{(i)} | P)$ for $i=0,1$. 
Then, from (\ref{Theta_max}) we have
\begin{align}
 \Theta(R_\alpha, q^{(\alpha)} | P) 
{\leq} 
\sum_{i=0,1} \alpha_i 
\Theta(R_i, q^{(i)} | P) {.}
\label{eq.25}
\end{align}
Therefore, 
\begin{align*}
&  G(R_\alpha, \Delta_\alpha | P) 
= 
\min_{
\scriptstyle 
q \in \mathcal{P(X\times Y)}: 
\atop
\scriptstyle 
{ {\rm E}_{q} [d(X,Y)] }
\leq \Delta_\alpha}
\Theta(R_\alpha, q | P) \\
&
\stackrel{\rm (a)}{\leq} 
\Theta(R_\alpha, q^{(\alpha)} | P) 
\stackrel{\rm (b)}{\leq} 
\sum_{i=0,1} \alpha_i 
\Theta(R_i, q^{(i)} | P) \\
&
\stackrel{\rm (c)}{=} 
\sum_{i=0,1} \alpha_i 
 G(R_i, \Delta_i | P). 
\end{align*}
Step (a) follows from (\ref{constraint_Delta_alpha}),
Step (b) follows from (\ref{eq.25}), and
Step (c) follows from (\ref{optimal_distribution}).

For the proof of Part c), the choice of 
$q_X=P$ gives $G(R,\Delta|P)=0$, 
if $R\geq R(\Delta|P)$. 
If $R<R(\Delta | P)$, 
the choice of $q_X=P$ makes 
the first term of the objective
function strictly positive, while 
any choice of $q \neq P$, 
$D(q || P)$ is strictly positive. 
This completes the proof of Part c).

For the proof of Part d),
let $q^*$ be a joint distribution that attains
$G(R', \Delta |P)$.
Then,
\begin{align*}
 G(R, \Delta |P) 
\leq & 
| I(q_X^*, q_{Y|X}^*) - R |^+ + D(q_X^* || P)\\
\stackrel{\rm (a)}\leq & (R'-R) + 
| I(q_X^*, q_{Y|X}^*) - R' |^+ \\
& + D(q_X^* || P)\\
= &(R'-R) + G(R', \Delta|P). 
\end{align*}
Step (a) follows from $|x|^+ \leq |x-c|^+ + c $ for $c\geq 0$.
This completes the proof. \hfill\IEEEQED

}
{
In the {following}, we give 
definitions of three functions that
are related to $G(R, \Delta|P)$ {and show their properties.} 
Then we give a lemma, from which  
a parametric expresssion of $G(R, \Delta|P)$
is derived.}

For $0\leq \lambda\leq 1, R\geq 0, \Delta \geq 0$ 
and $\mu\geq 0$,
we define 
\begin{align}
G^{(\lambda)} (R, \Delta|P) & \defeq  
\min_{
\scriptstyle 
q_{XY} \in {\cal P(X \times Y)}: \atop
\scriptstyle 
{ {\rm E}_{q_{XY}} [ d(X,Y) ] }
\leq \Delta 
} 
\{ \lambda ( I(q_X, q_{Y|X}) - R )  \notag \\
& \hspace{5mm} + D(q_X || P)  \}, 
\label{GCKlambda} \\
\Omega^{(\mu, \lambda)} (P) 
& \defeq
\min_{q_{XY} \in {\cal P(X \times Y)}} 
\{ 
\lambda I( q_X, q_{Y|X} )
\notag\\
& \hspace{5mm}
+ D(q_{X} || P ) + \mu 
{ {\rm E}_{q_{XY}} [ d(X,Y) ] }
\}, 
\\
G^{(\mu, \lambda)}(R, \Delta |P) 
&  \defeq 
\min_{q_{XY}  \in {\cal P(X \times Y)}} 
 \big\{ 
 \lambda (I(q_X, q_{Y|X}) - R )\notag\\
& \hspace{5mm} + D(q_X||P) - \mu (\Delta - 
{ {\rm E}_{q_{XY}} [d(X,Y)] }
)
\big\} \notag\\
& = \Omega^{(\mu, \lambda)} (P) - \lambda R - \mu \Delta. 
\label{G_CK_mu_lambda}
\end{align}
{
It is obvious from these definitions that the joint disribution
$q_{XY}$ that minimizes $\Omega^{(\mu, \lambda)}$
also minimizes $G^{(\mu, \lambda)}(R, \Delta|P)$ irrespective of
the values of $R$ and $\Delta$.
$G^{(\lambda)}(R, \Delta | P)$ will be used in Section V for
calculating the cut-off rate. 
}

\newcommand{\PrIII}{
The function $G^{(\lambda)}(R, \Delta|P)$
satisfies the following property: \

\newpage

\begin{pr}
\label{pr:G_CK_lambda}
\

\begin{itemize}
\item[a)] 
$G^{(\lambda)}(R, \Delta|P)$ is a monotone decreasing
function of $R \geq 0$
for a fixed $\Delta \geq 0$ 
and is a monotone decreasing function of $\Delta \geq 0$
for a fixed $R\geq 0$.
\item[b)] $G^{(\lambda)}(R, \Delta|P)$ is a convex function of $(R,\Delta)$.
\end{itemize}
\end{pr}

}

\PrIII

\conforfull{\ref{proofPrIII}}
%
\newcommand{\proofPrIII}{
\subsection{Proof of Property \ref{pr:G_CK_lambda}} 
\label{proofPrIII}
In this appendix we prove Property~\ref{pr:G_CK_lambda}.
By definition, Part a) is obvious.
For the proof of Part b), 
let $q^{(0)}$ and $q^{(1)}$ be 
joint distribution functions that attain 
$G^{(\lambda)}(R_0, \Delta_0|P)$ and 
$G^{(\lambda)}(R_1, \Delta_1|P)$,
respectively. 
Denote 
\begin{align}
&\Theta^{(\lambda)} (R, q | P) 
\defeq  \lambda [ I(q_X, q_{Y|X}) - R ] + D(q_X || P).
\label{Theta_lambda}
\end{align}
By definition, we have
\begin{align}
G^{(\lambda)} (R_i, \Delta_i |P) 
= 
\Theta^{(\lambda)} (R_i, q^{(i)} |P)
\mbox{ for } i = 0, 1.
\label{optimal_distribution_lambda}
\end{align}
For $\alpha_1 = \alpha \in [0,1]$ and 
$\alpha_0 = 1-\alpha$,
we set 
$R_\alpha = \alpha_0 R_0 + \alpha_1 R_1$, 
$\Delta_\alpha = \alpha_0 \Delta_0 + \alpha_1 \Delta_1$, 
and 
$q^{(\alpha)} = \alpha_0 q^{(0)} + \alpha_1 q^{(1)}$.
By linearity of 
$ 
{ {\rm E}_q[d(X,Y)] }
$ with respect to $q$,
we have that
\begin{align}
{ {\rm E}_{q^{(\alpha)} } [ d(X, Y) ] }
= \sum_{i=0,1} \alpha_i 
{ {\rm E}_{q^{(i)}} [ d(X, Y) ] }
\leq \Delta_{\alpha} . 
\label{constraint_Delta_alpha_lambda}
\end{align}
{Since}
\begin{align*}
& \lambda I(q_X, q_{Y|X}) + D(q_X||P)  \\
& = 
\lambda [ I(q_X, q_{Y|X}) + D(q_X||P) ]
+ (1 - \lambda) D(q_X||P) \\
& = 
\lambda \sum_{x,y} q_{XY}(x,y)
\log 
\frac{q_{X|Y}(x|y)}{P(x)}
+ (1-\lambda) D(q_X||P)
\end{align*}
is convex with
respect to $q_{XY}$, 
we have
\begin{align}
& \Theta^{(\lambda)}(R_\alpha, q^{(\alpha)} | P) \notag\\
& = \lambda [ I( q_X^{(\alpha)}, q_{Y|X}^{(\alpha)}) -R_\alpha ] 
+ D(q_X^{(\alpha)} || P) \notag\\
& \leq \sum_{i=0,1} \alpha_i  
\left\{
\lambda [ I( q_X^{(i)}, q_{Y|X}^{(i)}) - R_i ]
+ D(q_X^{(i)} || P)
\right\} \notag \\
& \stackrel{\rm (a)}{=} \sum_{i=0,1} \alpha_i 
\Theta^{(\lambda)} (R_i, q^{(i)} | P). 
\label{eq.24}
\end{align}
Step (a) follows from the definition
of $\Theta^{(\lambda)} (R_i, q^{(i)} | P)$ for $i=0,1$. 
Therefore, 
\begin{align*}
&  G^{(\lambda)} (R_\alpha, \Delta_\alpha | P) 
= 
\min_{
\scriptstyle 
q \in \mathcal{P(X\times Y)}: \atop
\scriptstyle 
{ {\rm E}_{q} [d(X,Y)] }
\leq \Delta_\alpha}
\Theta^{(\lambda)} (R_\alpha, q | P) \\
&
\stackrel{\rm (a)}{\leq} 
\Theta^{(\lambda)} (R_\alpha, q^{(\alpha)} | P) 
\stackrel{\rm (b)}{\leq} 
\sum_{i=0,1} \alpha_i 
\Theta^{(\lambda)} (R_i, q^{(i)} | P) \\
&
\stackrel{\rm (c)}{=} 
\sum_{i=0,1} \alpha_i 
 G^{(\lambda)} (R_i, \Delta_i | P). 
\end{align*}
Step (a) follows from (\ref{constraint_Delta_alpha_lambda}),
Step (b) follows from (\ref{eq.24}), and
Step (c) follows from (\ref{optimal_distribution_lambda}).
This completes the proof.\hfill\IEEEQED
}

%
Then, we have the following lemma:
\begin{lemma} \label{lm:G_CK_mu}
For any $R\geq 0, \Delta\geq 0$, we have
\begin{align}
G(R, \Delta|P) 
= \max_{0\leq \lambda \leq 0}
G^{(\lambda)}(R, \Delta|P). 
\label{Lemma2-1}
\end{align}
For any $0\leq \lambda \leq 1$, 
$R\geq 0$, and $\Delta \geq 0$,
we have
\begin{align}
 G^{(\lambda)}(R, \Delta | P) 
= \max_{\mu \geq 0} 
G^{(\mu, \lambda)}(R, \Delta|P).
\label{Gmulambda}
\end{align}
Eqs.(\ref{Lemma2-1}) and (\ref{Gmulambda}) imply that 
for {any} $R\geq 0$, $\Delta\geq 0$, we have 
\begin{align*}
G(R, \Delta | P)
=
\max_{0\leq \lambda \leq 1}
\max_{\mu \geq 0}
G^{(\mu, \lambda)}(R, \Delta |P). 
\end{align*}

\end{lemma}

\conforfull{\ref{ProoflmII}} Properties 1 and 2 are needed to
prove (\ref{Lemma2-1}) and  (\ref{Gmulambda}), respectively.


{
It follows from Lemma~\ref{lm:G_CK_mu} that 
$G(R, \Delta |P)$ is obtained by 
maximizing $G^{(\mu, \lambda)} (R, \Delta |P)$ with
respect to $(\mu, \lambda)$.
The first step for obtaining 
$G^{(\mu, \lambda)} (R, \Delta |P)$ is
to calculate the joint distribution that
minimizes $\Omega^{(\mu, \lambda)} (P)$.
In the next section, we give an algorithm to
obtain such a joint distribution.
}

\newcommand{\prooflmII}{
\subsection{Proof of Lemma \ref{lm:G_CK_mu}}
\label{ProoflmII}
In this appendix we prove Lemma \ref{lm:G_CK_mu}.
First, we prove Eq.(\ref{Lemma2-1}).
For any $\lambda \in [0,1]$, 
we have $|x|^+ \geq \lambda x$.
{Let $\hat q$ be a joint distribution 
that attains $G(R, \Delta | P)$. }
Then, we have 
\begin{align*}
G(R, \Delta | P) 
= & 
{
| I(\hat q_X,\hat q_{Y|X} ) - R |^+ + D(\hat q_X||P) 
}\\
\geq & 
{
\lambda [ I(\hat q_X,\hat q_{Y|X} ) - R ] + D(\hat q_X||P) 
}\\
\geq & 
\min_{
\scriptstyle 
q \in \mathcal{P(X\times Y)}: \atop
\scriptstyle 
{ {\rm E}_{q} [ d(X,Y) ] }
\leq \Delta 
}
\big\{
\lambda [ I(q_X, q_{Y|X} ) - R ] + D(q_X||P)
\big\} \\
= & G^{(\lambda)}(R, \Delta | P) {.}
\end{align*}
Thus, 
$$ 
G(R, \Delta | P) 
\geq 
\max_{0\leq \lambda \leq 1}
G^{(\lambda)}(R, \Delta | P). 
$$
Hence, it is sufficient to show that 
there exists a $\lambda \in [0,1]$
such that 
$ G(R, \Delta | P)
\leq 
G^{(\lambda)}(R, \Delta | P)
$.
From Property \ref{pr:G_CK}, 
there exists a $\lambda \in [0,1]$ such that
for any $R'\geq 0$ we have
\begin{align}
 G(R', \Delta | P)
\geq 
G(R, \Delta | P)
- \lambda (R'-R) {.}
\label{eq:G_CK}
\end{align}
Fix the above $\lambda$. 
Let $q^*$ be a joint distribution
that attains
$G^{(\lambda)}(R, \Delta | P)$.
Set $R' = I(q_X^*, q_{Y|X}^*)$.
Then we have
\begin{align}
& G(R, \Delta | P) \notag \\
& \stackrel{\rm (a)}\leq 
 G(R', \Delta | P) + \lambda (R'-R) \notag \\
&=  
\min_{
\scriptstyle 
q_{XY}: \atop 
\scriptstyle 
{ {\rm E}_{q} [d(X,Y)] }
\leq \Delta } 
 \big\{ 
 | I(q_X, q_{Y|X}) - R' |^+ + D(q_X||P) \big\} \notag \\
& \hspace{1cm } 
+ \lambda (R'-R) \notag \\
& \leq 
| I(q_X^*, q_{Y|X}^*) - R' |^+ + D(q_X^*||P) + \lambda (R'-R) \notag \\
& \stackrel{\rm (b)}= 
 D(q_X^*||P) + \lambda [ I(q_X^*, q_{Y|X}^*) - R ] \notag \\
&= 
 G^{(\lambda)}(R, \Delta | P). 
\end{align}
Step (a) follows from
(\ref{eq:G_CK}) and Step (b) comes
from the choice of $R' = I(q_X^*, q_{Y|X}^*)$.
Therefore, there exists a $0 \leq \lambda \leq 1$
such that 
$G^{(\mu)}(R, \Delta | P) = 
G^{(\mu, \lambda)}(R, \Delta | P) $. 

}


\newcommand{\prooflmIII}{
Next, we prove (\ref{Gmulambda}). 
From its formula, it is obvious that
$$
G^{(\lambda)}(R, \Delta|P) 
\geq \max_{\mu \geq 0}
G^{(\mu, \lambda)}(R, \Delta|P). 
$$
Hence, it is sufficient to show that
for any $R\geq 0$ and $\Delta\geq 0$, 
there exists $\mu\geq 0$ such that
\begin{align}
G^{(\lambda)}(R, \Delta|P) 
\leq 
G^{(\mu, \lambda)}(R, \Delta|P). 
\label{ineq_G_CK_mu_lambda}
\end{align}
From Property \ref{pr:G_CK_lambda} part a) and b), $G^{(\lambda)}(R, \Delta|P)$
is a monotone decreasing and convex function of $\Delta \geq 0$ 
{for a 
fixed $R$}. Thus,
there exists $\mu \geq 0$ such that for any $\Delta'\geq 0$,
the following inequality holds:
\begin{align}
  G^{(\lambda)}(R, \Delta'|P) \geq 
  G^{(\lambda)}(R, \Delta|P) 
  - \mu (\Delta' - \Delta) {.}
  \label{convexity_G_CK_lambda}
\end{align}
Fix the above $\mu$. Let $q^*$ be
a joint distribution that attains
$G^{(\mu, \lambda)}(R, \Delta|P)$.
Set $\Delta' = 
{ {\rm E}_{q^*} [ d(X,Y) ] }
$.
Then, we have 
\begin{align*}
& G^{(\lambda)}(R, \Delta|P) 
  \stackrel{\rm (a)}{\leq} 
  G^{(\lambda)}(R, \Delta'|P) 
  - \mu (\Delta - \Delta') \\
& {= 
  \min_{
\scriptstyle 
q: \atop 
\scriptstyle
{ {\rm E}_{q} [ d(X,Y) ] }
\leq \Delta'
}
\{ 
\lambda [ I(q_X, q_{Y|X}) - R ] + D(q_X||P) \} }\\
&\hspace{5mm} {- \mu (\Delta - \Delta'  ) }
\\
& \stackrel{\rm (b)}\leq 
  \lambda [ I(q_X^*, q_{Y|X}^*) - R ] + D(q_X^*||P)
  - \mu (\Delta - 
{ {\rm E}_{q^*} [ d(X,Y) ] }
) \\
& = G^{(\mu, \lambda)}(R, \Delta|P). 
\end{align*}
Step (a) follows from (\ref{convexity_G_CK_lambda})
and Step (b) follows from the definition of
$G^{{(\lambda)}}(R, \Delta'|P)$
and the choice of $\Delta' = 
{ {\rm E}_{q^*} [ d(X,Y) ] }
$. 
Thus, for any $\Delta\geq 0$, we have (\ref{ineq_G_CK_mu_lambda}) 
for some $\mu \geq 0$.
This completes the proof. \hfill\IEEEQED

}


\newcommand{\Gspmu}{

We introduce the following functions and show their properties:
For $\mu, R, \Delta \geq 0$, we define 
\begin{align*}
& G_{\rm sp}^{(\mu)}(R, \Delta |P) \\
&= 
\min_{
\scriptstyle 
 q \in \mathcal{P(X \times Y)}: 
\atop
\scriptstyle 
I(q_X, q_{Y|X}) \leq R
}
\big\{ 
D(q_X||P) 
 - \mu (\Delta - 
{ {\rm E}_{q} [d(X,Y)] }
)
\big\}.
\end{align*}

\begin{pr} \label{pr:G_CK_sp}
\
\begin{itemize}
\item[a)]
For a fixed $\Delta \geq 0$, 
$G_{\rm sp}^{(\mu)}(R, \Delta |P) $ is a monotone
decreasing function of $R\geq 0$.
For a fixed $R \geq 0$, 
$G_{\rm sp}^{(\mu)}(R, \Delta |P) $ is a monotone
decreasing function of $\Delta \geq 0$. 
\item[b)] For $\mu \geq 0$, $R \geq 0$, 
and $\Delta\geq 0$, we have 
$$
  G^{(\mu)}(R, \Delta | P ) = G_{\rm sp}^{(\mu)}(R, \Delta | P).
$$
\end{itemize}
\end{pr}

{\it Proof:}
Property \ref{pr:G_CK_sp} part a) is obvious. 
We prove the part b) here.
For $\mu, R, \Delta \geq 0$, define
\begin{align}
& \hat G^{(\mu)}(R, \Delta |P) 
= \min_{
\scriptstyle 
 q_{XY} \in \mathcal{P(X\times Y)}: 
\atop 
\scriptstyle 
I(q_X, q_{Y|X}) \geq R
}
\big\{ 
I(q_X, q_{Y|X}) - R  \notag\\
& \hspace{20mm} + D(q_X||P) 
- \mu (\Delta - 
{ {\rm E}_{q} [d(X,Y)] }
)
\big\}. \label{Ghat_mu}
\end{align}
It is obvious that 
\begin{align}
 G^{(\mu)}(R, \Delta |P) 
 = \min 
\{
\hat G^{(\mu)}(R, \Delta |P) , 
G_{\rm sp}^{(\mu)}(R, \Delta |P)\}. \label{G_CK_max} 
\end{align}
Since
\begin{align*}
& I(q_X, q_{Y|X}) + D(q_X || P)
+ \mu 
{ {\rm E}_{q}[ d(X,Y) ] }
\\ 
&= 
\sum_{y \in \mathcal{Y}}
q_{Y}(y) 
\sum_{x \in \mathcal{X}}
q_{X|Y}(x|y)
\log
\frac{
q_{X|Y} (x|y)
}
{P(x) \exp(-\mu d(x,y))
}
\end{align*}
is a linear function of $q_{Y}$, 
Eq.(\ref{Ghat_mu}) is attained by some $q_{Y}$ satisfying 
$I(q_X, q_{Y|X}) = R$. Thus, 
\begin{align*}
& \hat G^{(\mu)}(R, \Delta |P) \\
& = \min_{
\scriptstyle 
q_{XY} \in \mathcal{P(X \times Y)}: \atop
\scriptstyle 
I(q_X, q_{Y|X}) = R
}
 \big\{ D(q_X||P)
- \mu (\Delta - E_{q} [d(X,Y)] )
\big\}\\
& \geq G_{\rm sp} ^{(\mu)}(R, \Delta |P).
\end{align*}
Hence, from (\ref{G_CK_max}), 
we have $G^{(\mu)}(R, \Delta |P)
= G_{\rm sp}^{(\mu)}(R, \Delta |P)$, 
completing the proof.
\hfill \IEEEQED

}

\section{Distribution Updating Algorithm}
{In this section, 
we propose an iterative algorithm for computing 
$\Omega^{(\mu, \lambda)}(P)$. 
Computation of $G(R,\Delta|P)$ from 
$\Omega^{(\mu, \lambda)}(P)$ is straightforward.
We observe that 
\begin{align*}
& \Omega^{(\mu, \lambda)}(P) \\
& = \min_{q_{XY}} \{
\lambda I(q_X, q_{Y|X}) 
+ D(q_X || P)  + \mu 
{ {\rm E}_{q_{XY}} [ d(X, Y) ] }
\} \\
& = \min_{q_{XY}} {\rm E}_{q_{XY}}
\left[ 
\log 
\frac{q_X^{1-\lambda}(X) q_{X|Y}^{\lambda}(X|Y)
\exp ( \mu d(X,Y)) }
{P(X) }
\right].
\end{align*}
Thus, for computing $\Omega^{(\mu, \lambda)}(P)$, 
we should find 
a joint distribution that 
minimizes 
the expectation of 
$$
\omega_{q}^{(\mu, \lambda)}(x,y)
\defeq
\log 
\frac{q_X^{1-\lambda}(x) q_{X|Y}^{\lambda}(x|y)
\exp ( \mu d(x,y)) }
{P(x) } 
$$
with respect to $q_{XY}$.}
Let us define 
\begin{align*}
& F^{(\mu, \lambda)} (p,q)
 \defeq  
{\rm E}_{q} 
\left[
\omega_{p}^{(\mu, \lambda)}(X,Y)
\right]
+ D(q || p ), 
\end{align*}
where $p=p_{XY}$ and $q=q_{XY}$ are two probability
distributions taking values on $\mathcal{ P( X \times Y ) }$. 

We have the following two lemmas:
\begin{lemma} \label{lm_CK_p}
For a fixed $q$, $F^{(\mu, \lambda)}(p,q)$
is minimized by $p=q$ and its minimum value is
\begin{align*}
F^{(\mu, \lambda)}(q,q) 
& = {\rm E}_{q} [ \omega_q^{(\mu, \lambda)}(X,Y) ]
\\
& = 
\lambda I(q_X, q_{Y|X}) +
D(q_X||P)
+ \mu 
{ {\rm E}_{q} [ d(X,Y) ] }.
\end{align*}
This implies that 
\begin{align}
&  \min_{p,q} F^{(\mu, \lambda)}(p,q)
= \min_{q  } F^{(\mu, \lambda)}(q,q)\notag\\
&
 = \min_{q} {\rm E}_q \left[
\omega_{q}^{(\mu, \lambda)}(X,Y)
\right]
{=\Omega^{(\mu, \lambda)}(P)}.
\label{max_CK_1}
\end{align}
\end{lemma}

{\it Proof:}
We have 
\begin{align*}
& F^{(\mu, \lambda)} (p,q) \notag\\
= & 
{\rm E}_{q} 
\left[
\log
{\frac{p_X^{1-\lambda}(X) p_{X|Y}^\lambda(X|Y) }
{P(X) \exp( - \mu d(X,Y) )}}
\right]
+ {\rm E}_{q} 
\left[
\log 
\frac{q(X,Y)}{p(X,Y)}
\right]
\notag \\
= & 
{\rm E}_{q} 
\left[
\log
\frac{q_X^{1-\lambda}(X) q_{X|Y}^\lambda(X|Y) }
{P(X) \exp( - \mu d(X,Y) )}
\right]\notag\\
& + 
{\rm E}_{q} 
\left[
\log
{\frac{p_X^{1-\lambda}(X) p_{X|Y}^\lambda(X|Y)}
{q_X^{1-\lambda}(X) q_{X|Y}^\lambda(X|Y)}}
\frac{q(X,Y)}{p(X,Y)}
\right]
\notag \\
= & 
{ 
{\rm E}_{q} 
\left[
\omega_{q}^{(\mu, \lambda)}(X,Y) 
\right]  + 
{\rm E}_{q} 
\left[
\log
\frac
{q_{Y|X}^{1-\lambda}(Y|X) q_{Y}^\lambda(Y)}
{p_{Y|X}^{1-\lambda}(Y|X) p_{Y}^\lambda(Y)}
\right]
}
\notag\\
= & 
{\rm E}_{q} 
\left[
\omega_{q}^{(\mu, \lambda)}(X,Y)
\right] \notag\\
& + (1-\lambda) D(q_{Y|X}|| p_{Y|X})
 + \lambda D(q_{Y}||p_{Y}). 
\end{align*}
Hence, by non-negativity of divergence we have
\begin{align*}
F^{(\mu, \lambda)}(p,q) 
\geq & 
{\rm E}_{q} 
\left[
\omega_{q}^{(\mu, \lambda)}(X,Y)
\right],
\end{align*}
where equality holds if $p=q$.
This completes the proof. 
\hfill\IEEEQED


\begin{lemma} \label{lm_CK_q}
For a fixed $p$, $F^{(\mu, \lambda)}(p,q)$ 
is minimized by 
\begin{align*}
q(x, y) 
& = 
\frac{1}{\Lambda^{(\mu, \lambda)}_{p} }
\frac{P(x) \exp(-\mu d(x,y)) p_{XY}(x,y)}
{p_X^{1-\lambda}(x) p_{X|Y}^{\lambda}(x|y)}\\
& \defeq
\hat q (p) (x, y),
\end{align*}
where $\Lambda^{(\mu, \lambda)}_{p} $ is a normalization factor defined by
\begin{align*}
\Lambda^{(\mu, \lambda)}_{p}  
&= {\rm E}_{p} 
\left[
  \frac{ P(X) \exp(-\mu d(X,Y)) }
  { p_X^{1-\lambda}(X) p_{X|Y}^{\lambda}(X|Y) }
\right]\\
&= {\rm E}_{p} 
\left[
\exp \{ - \omega^{(\mu, \lambda)}_{p}(X,Y) \} 
\right]
\end{align*}
and its minimum value is 
\begin{align*}
  F^{(\mu, \lambda)}(p, \hat q(p)) 
&= - \log \Lambda^{(\mu, \lambda)}_{p} \\
&= - \log 
{\rm E}_p 
\left[
\exp \{ - \omega^{(\mu, \lambda)}_{p}(X,Y) \} 
\right].
\end{align*}
This implies that
\begin{align}
 &   \min_{p,q} F^{(\mu, \lambda)} (p,q) 
   = \min_{p  } F^{(\mu, \lambda)} (p, \hat q(p) ) \notag \\
 & = \min_{p  } \left( - \log 
{\rm E}_p 
\left[
\exp \{ - \omega^{(\mu, \lambda)}_{p}(X,Y) \} 
\right] \right) . \label{max_CK_2}
\end{align}

\end{lemma}

{\it Proof:}
{We have} 
\begin{align*}
& F^{(\mu, \lambda)} (p,q) \notag \\
= & 
{\rm E}_{q} \left[
\log 
\frac{p_X^{1-\lambda}(X) p_{X|Y}^{\lambda}(X|Y) \exp\{\mu d(X,Y)\}q(X,Y)}
{P(X)  p(X,Y)}
\right] \notag \\
= & 
{\rm E}_{q} \left[
\log 
\frac
{q(X,Y)}
{
  \frac{1}{\Lambda^{(\mu, \lambda)}_{p} } 
  \frac{ P(X) \exp(-\mu d(X,Y)) p(X,Y) }
  {p_X^{1-\lambda}(X) p_{X|Y}^{\lambda}(X|Y) }
}
\right] - \log \Lambda^{(\mu, \lambda)}_{p}  \notag \\
\geq & - \log \Lambda^{(\mu, \lambda)}_{p} {,} 
\end{align*}
where the last inequality comes from the non-negativity
of the divergence.
Equality holds if $q(x,y)= \hat q (p)(x, y)$.
This completes the proof. 
\hfill \IEEEQED

By Lemmas \ref{lm_CK_p} and \ref{lm_CK_q},
we can obtain an iterative algorithm
for computing $\Omega^{(\mu, \lambda)}( P )$ 
as follows: 

\noindent
\underline{Distribution updating algorithm}

\begin{itemize}
\item[ 1) ]
{Choose an initial 
probability vector
$q^{[1]}$ arbitrarily such that all its components are nonzero. }
\item[ 2) ]
Then, iterate the following steps for
$t=1,2,3,\ldots,$
\begin{align}
q^{[t+1]}(x,y) 
&= 
\frac{\exp \left\{ -\omega_{ q^{[t]} }^{(\mu, \lambda)} (x, y) \right\}
q^{[t]}(x, y)
}{ {\Lambda^{(\mu, \lambda)}_{q^{[t]}} }} \label{update_CK}\\
&\defeq { \hat q(  q^{[t]})(x,y)}, \notag
\end{align}
where 
$\displaystyle 
\Lambda_{q^{[t]} }^{(\mu, \lambda)}
= {\rm E}_{ q^{[t]} } \left[
\exp \left\{ - \omega_{ q^{[t]} }^{(\mu, \lambda)} (X, Y) \right\} 
\right]
.$

\end{itemize}
We have the following proposition:

\begin{proposition} \label{proposition:algorithm:monotone}
For $t={1,2,3,}\ldots$, we have
\begin{align*}
&\quad\:\: 
{F(q^{[1]}, q^{[1]}) 
\stackrel{\rm ({a})}{\geq} 
F(q^{[1]}, q^{[2]}) 
\stackrel{\rm ({b})}{\geq} 
F(q^{[2]}, q^{[2]}) 
{\geq} }
\cdots \notag\\
&\stackrel{\rm \:\:}{\geq} F(q^{[t]}, q^{[t]}) 
\notag\\
&
\stackrel{\rm ({a})}{\geq} F(q^{[t]}, q^{[t+1]})
= - \log {\rm E}_{q^{[t]}} 
\left[
\exp \left\{ - \omega_{ q^{[t]} }^{(\mu, \lambda)}(X,Y) \right\}      
\right] 
\notag\\
& \stackrel{\rm ({b})}{\geq} F(q^{[t+1]}, q^{[t+1]})
={\rm E}_{q^{[t+1]}} 
\left[
 \omega_{ q^{[t+1]} }^{(\mu, \lambda)}(X,Y)     
\right] 
\notag\\
& \stackrel{}{\geq} \cdots 
{\geq} 
\min_{q} \left\{
-\log {\rm E}_{q}
\left[
\exp \left\{ - \omega_{q}^{(\mu, \lambda)}(X,Y) \right\} 
\right] \right \}
\notag\\
&
\stackrel{\rm (c)}{=} 
\min_{q}{\rm E}_{q}\left[ \omega_{q}^{(\mu, \lambda)}(X,Y)\right]
=\Omega^{(\mu, \lambda)}(P).
\end{align*}
\end{proposition}

{\it Proof:} 
{
Step (a) follows from Lemma {\ref{lm_CK_q}}. 
Step (b) follows from Lemma {\ref{lm_CK_p}}.
Step (c) follows from 
Eq.~(\ref{max_CK_1}) in Lemma \ref{lm_CK_p} and 
Eq.~(\ref{max_CK_2}) in Lemma \ref{lm_CK_q}. 
This completes the proof.
\hfill\IEEEQED}

\section{ Convergence of the algorithm }
Proposition \ref{proposition:algorithm:monotone} shows that 
$F(q^{[t]}, q^{[t]})$ {decreases} by updating
the probability distribution $q^{[t]}$ 
according to (\ref{update_CK}). 
This section shows that $q^{[t]}$ converges to the optimal 
distribution. 
We have the following theorem:

\begin{theorem} \label{theorem:convergence}
For any $0\leq \lambda \leq 1$ and any $\mu \geq 0$
probability vector $q^{[t]}$ defined by (\ref{update_CK})
converges to the optimal distribution $q^*$ that attains
the minimum of 
\begin{align*}
{\rm E}_{q} \left[
\omega_{q}^{(\mu, \lambda)} (X, Y)
\right]
&= 
\lambda I(q_X, q_{Y|X}) + D(q_X||P) \\
& \phantom{=} + \mu 
{{\rm E}_{q} [d(X,Y)] }
\end{align*}
in the definition of $\Omega^{(\mu, \lambda)}(P)$.
\end{theorem}

{\it Proof:}
{By definition, we have}
$
{\rm E}_{q^*} [  \omega_{q^*}^{ (\mu, \lambda) } (X,Y) ]
= 
\Omega^{ (\mu, \lambda) } (P) 
$
{
and
$F(q^{[t]},q^{[t+1]}) = - \log \Lambda_{{q}^{[t]}}^{(\mu, \lambda)}$.} 
From Eq.(\ref{update_CK}), we have 
\begin{align}
\Lambda_{q^{[t]} }^{ (\mu, \lambda) } 
= \frac{q^{ [t] }(x, y)}{q^{ [t+1] }(x, y)} 
\exp \{ 
- \omega_{q^{ [t] }}^{ (\mu, \lambda) } (x, y) \}.
\label{trick_CK}
\end{align}
Hence,
{
\begin{align*}
& - \log \Lambda_{q^{[t]} }^{(\mu, \lambda)} 
  - \Omega^{(\mu, \lambda)} (P) 
\notag \\
= & 
- \mathrm{E}_{q^*} [\log \Lambda_{ q^{[t]} }^{(\mu, \lambda)}]
- \mathrm{E}_{q^*} [
\omega_{q^*}^{(\mu, \lambda)}(X, Y)]
\notag\\
\stackrel{\rm (a)}{=} & \mathrm{E}_{q^*} \left[ 
\log \frac{ q^{ [t + 1] }(X, Y) }{ q^{ [t] }(X, Y)}
+ \omega_{ q^{ [t] } }^{(\mu, \lambda)} (X, Y) 
- 
\omega_{ q^* }^{(\mu, \lambda)} (X, Y) 
\right] \notag\\
= & \mathrm{E}_{q^*}  
\Bigg[
\log \frac{ q^{ [t + 1] }(X, Y) }{ q^{ [t] }(X, Y)}
+  
\log \left \{ 
  \frac{q_{{X}}^{[t]}(X)}{q_{{X}}^*(X)}
  \right\}^{1-\lambda} \notag\\
& + \log
  \left\{ \frac{q_{{X|Y}}^{[t]}(X|Y)}{q_{{X|Y}}^*(X|Y)} \right\}^{\lambda}
\Bigg] \notag\\
= & \mathrm{E}_{q^*} 
\left[
  \log \frac{ q^{ [t + 1] }(X, Y) }{ q^{ [t] }(X, Y)}
\right]
- (1-\lambda) D(q^*_{X} || q_{X}^{ [t] } ) \notag\\
&- \lambda  D(q^*_{X|Y} || q_{X|Y}^{ [t] } | q_{Y}^* ) 
\leq 
\mathrm{E}_{q^*}  
\left[
\log \frac{ q^{ [t + 1] }(X, Y) }{ q^{ [t] }(X, Y)}
\right],
\end{align*}
}
{where equality (a) holds because
Eq.(\ref{trick_CK}) holds for all $(x,y)\in \mathcal{X \times Y}$}. 
Thus,
\begin{align*}
0 \leq 
- \log \Lambda_{ q^{[t]} }^{ ({\mu,} \lambda) }
- \Omega^{ (\mu, \lambda) } (P) 
& \leq 
\mathrm{E}_{q^*}  
\left[
 \log 
\frac{ q^{ [t + 1] }(X, Y) }{ q^{ [t] }(X, Y)} 
\right]
\notag\\
& = D(q^* || q^{ [t] }) - D(q^* || q^{ [t + 1] }). 
\end{align*}
Therefore, we have 
$ D(q^* || q^{ [t] }) \geq D(q^* || q^{ [t + 1] }) $, 
which implies that the 
KL distance between 
$q^{[t]}$ and $q^*$ decreases by updating $q^{[t]}$.
Put $ 
- \log \Lambda_{ q^{[t]} }^{ (\mu, \lambda) }
- \Omega^{ (\mu, \lambda) } (P) 
= \xi_t$. Then 
{
\begin{align}
0 \leq \sum_{t=1}^T \xi_t& =D(q^* || q^{ [1] }) - D(q^* || q^{ [ T + 1] })
\notag\\ 
 & < D(q^* || q^{ [1] }). 
\label{ConvergeA_CK}
\end{align}
{$D(q^* || q^{ [1] })$ is finite because 
all components of $q^{[1]}$ are nonzero.}
By Proposition \ref{proposition:algorithm:monotone}, $\{\xi_t\}_{t\geq 1}$ 
is a monotone decreasing sequence. Then from Eq.(\ref{ConvergeA_CK}), 
we have 
$
0\leq T \xi_T \leq D(q^* || q^{ [1] }),
$
from which we have
$$
0\leq \xi_T\leq \frac{D(q^* || q^{ [1] })}{T}\to 0, \quad T\to\infty. 
$$
}
Hence, we have 
\begin{align*}
  \lim_{t \to \infty} 
\left\{ - \log \Lambda_{ q^{[t]} }^{(\mu, \lambda)} \right\}
= \Omega^{(\mu, \lambda)} (P),
\end{align*}
completing the proof.
\hfill$\IEEEQED$

{
As a corollary of Proposition \ref{proposition:algorithm:monotone}
and Theorem \ref{theorem:convergence},
we have the following result, 
which provides a new parametric expression
of $G^*(R, \Delta|P)=G(R, \Delta|P)$. 
\begin{corollary} 
  \label{corol:2}
\begin{eqnarray} 
& & G^{(\mu, \lambda)}
(R, \Delta|P) = - \lambda R - \mu \Delta \notag\\
& & \hspace{12mm} + \min_{p} \left\{ - \log {\rm E}_{p}
\left[
\frac{ P(X) \exp(-\mu d(X,Y)) }
{p_{X}^{1-\lambda}(X) p_{Y|X}^{\lambda}(Y|X)}
\right] \right\},
\nonumber\\
& &G^*(R, \Delta|P)=G(R, \Delta|P) \nonumber\\
& &= \max_{0 \leq \lambda \leq 1} 
\max_{\mu \geq 0} 
G^{(\mu, \lambda)}(R, \Delta|P)
\nonumber\\
& & = \max_{0\leq \lambda \leq 1} 
\max_{\mu \geq 0}  \min_{p}
\Bigg\{ - \lambda R - \mu \Delta \notag\\
& & \hspace{15mm} - \log {\rm E}_{p}
\left[
\frac{ P(X) \exp ( -\mu d(X,Y)) }
{ p_{X}^{1-\lambda}(X) p_{Y|X}^{\lambda}(Y|X)}
\right]
\Bigg\}. \nonumber
\end{eqnarray} 
\end{corollary}
}

{
The proposed algorithm calculates Csisz\'ar and K\"orner's exponent
that expresses the optimal exponent of {\it correct decoding
probability} for $R<R(\Delta|P)$, while 
Arimoto~\cite{Arimoto1976} has presented an iterative algorithm 
for computing an exponent function of {\it 
error probability}\footnote{{
Blahut~\cite{Blahut1974} gave a lower bound of the exponent 
function of error probability 
for $R>R(\Delta|P)$. 
{The optimal} exponent of the error probability 
for $R>R(\Delta|P)$ was determined by Marton~\cite{Marton1974}. 
}
}
derived by Blahut~\cite{Blahut1974}.  
In Arimoto's algorithm, output distribution $q_Y \in \mathcal{P(Y)}$
and conditional probability distribution $q_{Y|X} \in \mathcal{P(Y|X)}$
 are alternately updated. 
Unlike Arimoto's algorithm, 
a joint distribution over the input and output alphabets 
is updated iteratively in the proposed method.
Unfortunately, the proposed algorithm
cannot be directly applied to the computations of 
the exponent function of {\it error probability}
because 
{they involve} {mini-max} structure, {\it i.e, }
maximization with respect to stochastic matrices
and minimization with respect to input distribution. 
}


\section{ Computation of Cutoff Rate and the Rate Distortion Function }
\begin{figure}[t]
\centering
\includegraphics[width=7cm]
{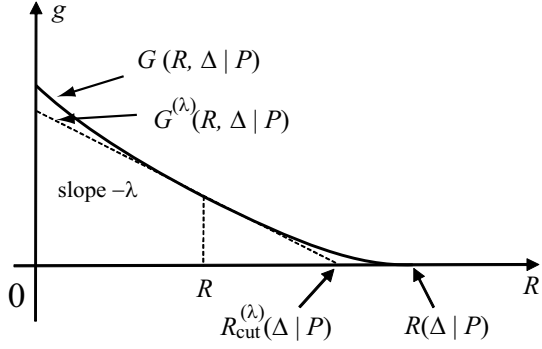}
\caption{{
The bold curve shows the exponent
function $G(R, \Delta|P)$ and the dashed line
shows the supporting line of slope $-1 \leq - \lambda \leq 0$
to the curve $G(R, \Delta|P)$.
${R_{\rm cut}^{(\lambda)}(\Delta|P)}$ is the 
$R$-axis intercept of the supporting line,
which {approaches} $R(\Delta|P)$ as $\lambda \to 0+$.
}
}
\label{cut_off_rate_CK}
\end{figure}
The proposed algorithm can be used for computing 
cutoff rate and the rate distortion function. 
First, we give the definition of the cutoff rate 
for lossy source coding. 
From (\ref{GCKlambda}), 
for a strictly positive $\lambda$, 
we have 
\begin{align}
& G^{(\lambda)}(R, \Delta|P) 
= -\lambda R \notag \\
& \hspace{0.5cm}
 + \lambda 
\min_{
\scriptstyle 
q_{XY} {\in \mathcal{P(X\times Y)}}: \atop
\scriptstyle 
{ {\rm E}_{q}[d(X,Y)] }
\leq \Delta }
\{
I(q_X, q_{Y|X}) + \frac1{\lambda} D(q_X||P)
\}. 
\label{Glambda}
\end{align}
%
%
%
%
%
{For fixed $\Delta \geq 0$ and $\lambda>0$, 
the right hand side of Eq.(\ref{Glambda}) is viewed as a linear function of $R$.
Moreover, from Eq.(\ref{Lemma2-1}) in Lemma \ref{lm:G_CK_mu}, }
$G^{(\lambda)}(R, \Delta|P)$ can be viewed as a
supporting line to the curve $G(R, \Delta|P)$ {with slope $-\lambda$}.
A rough sketch of the graph $g = G(R, \Delta|P)$ and
$g = G^{(\lambda)}(R, \Delta|P)$ is illustrated in
Fig.~\ref{cut_off_rate_CK}. 
From Property \ref{pr:G_CK}, $G(R, \Delta |P)$ takes 
positive value when $R<R(\Delta |P)$.
The cutoff rate is defined as $R$ that satisfies
$G^{(\lambda)}(R, \Delta|P)  = 0$, {\it i.e.,}
\begin{align}
& R_{\rm cut}^{(\lambda)} (\Delta|P) \notag \\
& \defeq  
\min_{\scriptstyle 
q_{XY} \in \mathcal{P(X \times Y)} : \atop
\scriptstyle 
{ {\rm E}_{q}[d(X,Y)] }
\leq \Delta }
\big\{
I(q_X, q_{Y|X}) + \frac{1}{\lambda } D(q_X||P)
\big\}. 
\label{def:CutOffRate}
\end{align}



{
The cutoff rate is calculated by using the proposed method as follows:
From {Eq.(\ref{Gmulambda}) in }Lemma \ref{lm:G_CK_mu}, we have
\begin{align}
& G^{(\lambda)} (R, \Delta | P) 
= 
\max_{\mu \geq 0}
G^{(\mu, \lambda)} (R, \Delta | P) \notag \\
& \hspace{1cm}= 
- \lambda R 
+ \lambda 
\max_{\mu \geq 0}
\left\{
\frac{1}{\lambda}\Omega^{(\mu, \lambda)}(P)
-\frac{\mu}{\lambda} \Delta 
\right\}.
\label{Glambda2}
\end{align}
From Eqs.~(\ref{Glambda}), (\ref{def:CutOffRate}), 
and (\ref{Glambda2}),
we have
\begin{align*}
R_{\rm cut}^{(\lambda)}(\Delta | P) =
\max_{\mu \geq 0}
\left\{
\frac{1}{\lambda}\Omega^{(\mu, \lambda)}(P)
-\frac{\mu}{\lambda} \Delta 
\right\}. 
\end{align*}
By Theorem 3, we can calculate 
$\Omega^{(\mu, \lambda)}(P)$ by the proposed algorithm.
Then, the cutoff rate is obtained by calculating
the maximum of  
$
\frac{1}{\lambda}\Omega^{(\mu, \lambda)}(P)
-\frac{\mu}{\lambda} \Delta$ with respect to $\mu\geq 0$. 
}


We show that 
$R_{\rm cut}^{(\lambda)} (\Delta|P)$ approaches 
$R(\Delta |P)$ as $\lambda \to 0+$. 
Let 
$\alpha = \min \{ \log |\mathcal{X}|, \log |\mathcal{Y}| \}
$ and 
$d_{\max} 
= \max_{( x, y ) \in ( \mathcal{X}, \mathcal{Y} ) } 
d(x,y)$.
We have the following {proposition}.

\begin{proposition}\label{proposition:cutoffrate}
$R_{\rm cut}^{(\lambda)}(\Delta |P)$ 
is a monotone decreasing function
of $\lambda$. 
Moreover, if $ \lambda \leq \frac{1}{8 \alpha}$, 
we have 
\begin{align}
0 
&\leq 
R (\Delta | P) -
R_{\rm cut} ^{(\lambda)} ( \Delta | P )  \leq 
c_1 \sqrt{\lambda} 
\left( \log \lambda^{-1} + c_2 \right), 
\label{eq:cutoff}
\end{align}
where 
$c_1 = \frac32 \sqrt{2 \alpha}$, 
$c_2 = 
\frac43 \log (|\mathcal{X}| |\mathcal{Y}|) 
- \log (2\alpha) 
+
\frac23 d_{\max} |R'(\Delta|P)|$, 
and 
$R'(\Delta|P) = \frac{\rm d}{{\rm d} \Delta} R(\Delta |P)$. 
This inequality implies that 
$$\lim_{ \lambda \to 0+ }
R_{\rm cut} ^{(\lambda)} ( \Delta | P )
=
R (\Delta | P).
$$
\end{proposition}

\conforfull{\ref{proof:proposition:cutoffrate}}

This proposition implies that by choosing a sufficiently 
small $\lambda>0$, we can use $R_{\rm cut}^{(\lambda)}(\Delta |P)$
as a good approximation of $R(\Delta | P)$ 
{for which} {accuracy is guaranteed by}
(\ref{eq:cutoff}).

\newcommand{\proofPropCutoffrate}{
\subsection{Proof of Proposition~\ref{proposition:cutoffrate}}
\label{proof:proposition:cutoffrate}
In this appendix, we prove 
Proposition~\ref{proposition:cutoffrate}. 
We {begin with} the following lemma:
\begin{lemma} 
If two probability distributions $p$ and $q$ on $\mathcal{X}$
satisfy 
$D(p||q) \leq \nu$ for a constant $\nu \leq \frac{1}{8}$,
we have
$$
| H(p) - H(q) | \leq \sqrt{ 2\nu} \log \frac{|\mathcal{X}|}{\sqrt{2\nu}} .
$$
\end{lemma}

{\it Proof:} 
From Pinsker's inequality, we have 
$$
D(p||q) \geq \frac12
\lVert p - q\rVert_1,
$$
where 
$\lVert p - q\rVert_1
= \sum_{\ x \in \mathcal{X}} | p(x) - q(x) |$.
It follows from Lemma 2.7 in~\cite{Csiszar-KornerBook}
that if $\lVert p - q\rVert_1 = \Theta \leq \frac12$, then we have 
$$
| H(p) - H(q) | \leq \Theta \log \frac{|\mathcal{X}|} {\Theta}.
$$
The lemma is proved by 
combining these two inequalities together with 
monotone increasing property of 
$\Theta \log \frac{|\mathcal{X}|}{\Theta}$ 
for $0\leq \Theta \leq \mathcal{|X|}/{\rm e}$. 
\hfill$\IEEEQED$

\

{\it Proof of Proposition 2:} First we show the monotonicity of 
$R_{\rm cut}^{(\lambda)}(R, \Delta |P)$
with respect to $\lambda$. Let $0<\lambda \leq \lambda' \leq 1$
and $q^*$ be a joint distribution that attains
$R_{\rm cut}^{(\lambda)}(R, \Delta |P)$.
Then, we have
\begin{align*}
R_{\rm cut}^{(\lambda')}(R, \Delta |P)
& \stackrel{\rm (a)}{\leq}
I(q^*_X, q_{Y|X}^*) + 
\frac1{\lambda'} D(q_X^* ||P) \\
& \stackrel{\rm (b)}\leq 
I(q^*_X, q_{Y|X}^*) + 
\frac1{\lambda} D(q_X^* ||P) \\
& = R_{\rm cut}^{(\lambda)}(R, \Delta |P).
\end{align*}
Step (a) follows from the definition and
step (b) follows from $\lambda \leq \lambda'$.

Next, we prove (\ref{eq:cutoff}). 
Let ${V}^*$ be a distribution on 
{$\mathcal{Y}$ given $\mathcal{X}$} 
that attains
$R(\Delta | P)$. Then, the choice of 
$(q_X, q_{Y|X} ) = (P, {V}^*)$ gives
\begin{align}
R_{\rm cut}^{(\lambda)} ( \Delta |P) 
\leq 
I(P, {V}^*) 
=
R(\Delta | P).
\label{Rcut_leq_IPVstar}
\end{align}
This gives the first inequality in (\ref{eq:cutoff}).

For the proof of second inequality {in (\ref{eq:cutoff})}, 
we first give an lower bound of $R_{\rm cut}^{(\lambda)}(\Delta|P)$
and then give an upper bound of $R(\Delta|P)$.
Let $q^*$ be
a joint distribution that attains $R_{\rm cut}^{(\lambda)}(R, \Delta|P)$.
Then, we have 
\begin{align*}
R_{\rm cut}^{(\lambda)}(\Delta | P )
= 
I(q^*_X, q_{Y|X}^*) + 
\frac1{{\lambda}} D(q_X^* ||P) .
\end{align*}
By the non-negativity of divergence, we have
\begin{align}
R_{\rm cut}^{(\lambda)}(\Delta | P )
\geq
I(q^*_X, q_{Y|X}^*) .
\label{Rcut_lowerbound1}
\end{align}
{By the non-negativity of mutual information, we also have
\begin{align*}
\frac1{{\lambda}} 
D(q_X^* ||P) 
& \leq 
R_{\rm cut} ^{(\lambda)}(\Delta | P ) \\
& \stackrel{\rm (a)}{\leq} 
I(P,{V}^*) 
\leq \min \{ \log |\mathcal{X}|, {\log} |\mathcal{Y}| \}.
\end{align*}
Step (a) follows from (\ref{Rcut_leq_IPVstar}).
}
Let $\alpha = \min \{ \log |\mathcal{X}|, {\log} |\mathcal{Y}| \}$.
Then, 
\begin{align}
D(q_X^* ||P) \leq \alpha \lambda . 
\label{Divergence_q_Xstar_and_P}
\end{align} 
Thus, 
$D(q_X^* ||P) \to 0$ as $\lambda \to 0+$, which shows 
$q_X^*$ converges to $P$. 
We have 
\begin{align}
& I(P, q_{Y|X}^*) - I(q_{ X}^*, q_{Y|X}^*) \notag \\
= & 
H(P) + H( Pq_{Y|X}^*) - H( ( P, q_{Y|X}^*) ) \notag \\
& -\{
H(q_X^* ) + H( q_{Y}^*) - H( q_{XY}^*) 
\} \notag \\
\leq & 
  |H(P) - H(q_X^* )|
+ |H( Pq_{Y|X}^*)- H( q_{Y}^*) | \notag \\
& + |H( ( P, q_{Y|X}^*) ) - H( q_{XY}^*) |. \label{eq.32}
\end{align}

From Lemma 5 and (\ref{Divergence_q_Xstar_and_P}), 
the first term of (\ref{eq.32})
is upper bounded by
$\sqrt{2 \alpha \lambda} \log \frac{|\mathcal{X}|}{\sqrt{2 \alpha \lambda}}
$ if $ \alpha \lambda \leq \frac{1}{8}$. 
By the chain rule of the divergence, we have 
$D(q_{XY}^* || (P, q_{Y|X}^*) ) = 
D(q_X^* || P ) 
+ D(q_{Y|X}^* || q_{Y|X}^* | q_X^* ) 
= D(q_X^* || P ) $ 
and 
$D( Pq_{Y|X}^* || q_{Y}^*) 
\leq D(q_{XY}^* || (P, q_{Y|X}^*) )
$. 
Thus, the second and the third terms of (\ref{eq.32})
are upper bounded by 
$\sqrt{2 \alpha \lambda} \log \frac{ \mathcal{ |Y| }}{\sqrt{2 \alpha \lambda}}$
and
$\sqrt{2 \alpha \lambda} \log \frac{\mathcal{ |X| |Y| }}{\sqrt{2 \alpha \lambda}}$, 
respectively. 
Therefore, by (\ref{Rcut_lowerbound1}) and (\ref{eq.32}), 
if $\lambda \leq \frac1{8\alpha}$, we have
\begin{align}
R_{\rm cut}^{(\lambda)}(\Delta|P) \geq & 
I(q^*, q_{Y|X}^{*}) \notag \\
\geq &I(P, q_{Y|X}^*) - 
\sqrt{2 \alpha \lambda} \log \frac{|\mathcal{X}|}{\sqrt{2 \alpha \lambda}} 
\notag \\
&\hspace{0.5cm}-
\sqrt{2 \alpha \lambda} \log \frac{|\mathcal{Y}|}{\sqrt{2 \alpha \lambda}} - 
\sqrt{2 \alpha \lambda} \log \frac{\mathcal{ |X| |Y| }}{\sqrt{2 \alpha \lambda}} \notag \\
=& I(P, q_{Y|X}^*) - 
\sqrt{2 \alpha \lambda} \log \frac{|\mathcal{X}|^2 |\mathcal{Y}|^2}{ (2 \alpha \lambda)^{{\frac32} } } .
\label{Upperbound_1}
\end{align}

Next, we give an upper bound of $R(\Delta|P)$. 
By the convexity of $R(\Delta|P)$, for any $\nu$, we have 
\begin{align}
R(\Delta + \nu) \geq R(\Delta | P) + \nu R'(\Delta|P),
\label{RateDistortionDerivative}
\end{align}
where
$$
R'(\Delta | P ) = \frac{\rm d}{{\rm d} \Delta} R(\Delta | P ).
$$
Note that $R(\Delta | P )$ is monotone decreasing function
of $\Delta$ and thus $R'(\Delta | P )\leq 0$. We have 
\begin{align*}
& { 
{\rm E}_{(P, q_{Y|X}^*)} [d(X,Y)] 
-
{\rm E}_{q^*} [d(X,Y)] 
} \\
&= \sum_{x \in \mathcal{X} } [ P(x) - q_X^*(x) ] \sum_{y \in \mathcal{Y} } 
q_{Y|X}^*(y|x) d(x,y)\\
& \leq \sum_{x \in \mathcal{X} } | P(x) - q_{X}^*(x) | d_{\max} 
=\lVert P- q_X^* \rVert_1 d_{\max}{.} 
\end{align*}
Therefore, the following inequality holds:
\begin{align}
{ {\rm E}_{(P, q_{Y|X}^*)} [d(X,Y)] 
}
&\leq 
{ {\rm E}_{q^*} [d(X,Y)] }
+ \lVert P- q_X^*\rVert_1 d_{\max}\notag\\
& \stackrel{\rm (a)}{\leq }
{ {\rm E}_{q^*} [d(X,Y)] }
+ \sqrt{2\alpha \lambda} d_{\max} \notag \\
&\stackrel{\rm (b)}{\leq}
\Delta + \sqrt{2\alpha \lambda} d_{\max}.
\label{distortion_of_qstar}
\end{align}
Step (a) follows from  (\ref{Divergence_q_Xstar_and_P}) and 
Pinsker's inequality.
Step (b) follows from the definition of $q^*$. 
Then, we have
\begin{align}
R(\Delta + \sqrt{2\alpha \lambda} d_{\max} | P ) 
&= \min_{\scriptstyle 
W \in \mathcal{P(Y|X)}: \atop
\scriptstyle 
{
{\rm E}_{(P,W)} [d(X,Y)] 
}
\leq \Delta + \sqrt{2\alpha \lambda} d_{\max} } 
I(P, W)
\notag \\
& \stackrel{\rm (a)}{ \leq} I( P, q_{Y|X}^* ).
\label{ineq36}
\end{align}
Step (a) follows from (\ref{distortion_of_qstar}). 
Then, we have the following inequality:
\begin{align}
R(\Delta|P) 
& \stackrel{\rm (a)}{\leq} 
  R(\Delta + \sqrt{2\alpha \lambda} d_{\max}|P) - \sqrt{2\alpha \lambda} d_{\max}R'(\Delta|P) \notag \\
&\stackrel{\rm (b)}{\leq} 
  I(P, q_{Y|X}^*) + \sqrt{2\alpha \lambda} d_{\max} |R'(\Delta|P)|.
\label{ineq37}
\end{align}
Step (a) follows from (\ref{RateDistortionDerivative}) 
with $\nu = \sqrt{2\alpha \lambda} d_{\max}$.
Step (b) follows from (\ref{ineq36}).

Then, we have 
the following: 
\begin{align*}
& R(\Delta | P) - R_{\rm cut}^{(\lambda)}(\Delta|P)\\
& { \stackrel{\rm (a)}{\leq }
 I(P, q_{Y|X}^*) + \sqrt{2\alpha \lambda} d_{\max} |R'(\Delta|P)| }\notag\\
& {\hspace{0.5cm} - \{ I(P, q_{Y|X}^*) - 
\sqrt{2 \alpha \lambda} \log \frac{|\mathcal{X}|^2 |\mathcal{Y}|^2}{ (2 \alpha \lambda)^{{\frac32} } }\} }\\
& = \sqrt{2 \alpha \lambda } \left\{ 
\log \frac{|\mathcal{X}|^2 |\mathcal{Y}|^2}{ (2 \alpha \lambda)^{ {\frac32}} }
+ d_{\max} |R'(\Delta|P)| \right\}\\
& {= \sqrt{2 \alpha \lambda } \left\{ 
- \frac32
\log \lambda
+ \log \frac{|\mathcal{X}|^2 |\mathcal{Y}|^2}{(2\alpha)^{\frac32}}
+ d_{\max} |R'(\Delta|P)| \right\} }\\
& = c_1 \sqrt{\lambda} ( \log \lambda^{-1} + c_2 ), 
\end{align*}
where 
$c_1 = \frac32 \sqrt{2 \alpha}$, 
$c_2 = 
\frac43 \log (|\mathcal{X}| |\mathcal{Y}|) 
- \log (2\alpha) 
+
\frac23 d_{\max} |R'(\Delta|P)|$. 
Step (a) follows from (\ref{Upperbound_1}) and (\ref{ineq37}).
{This completes the proof.} \hfill\IEEEQED

}

\newcommand{\conclusion}{
\section{Conclusion}
We have proposed an iterative algorithm for computing 
Csisz\'ar and K\"orner's exponent~\cite{Csiszar-KornerBook}
that expresses the optimal exponent
of correct decoding probability 
in lossy source coding when a rate $R$
is below the rate distortion function $R(\Delta | P)$. 
The proposed algorithm has a structure similar to
the one proposed by the authors~\cite{OohamaJitsumatsuISIT2015}
that computes Dueck and K\"orner's exponent 
in channel coding when the rate is above the capacity.
%
%
We have proven the joint distribution calculated by the proposed
algorithm converges to the optimal distribution that achieves
Csisz\'ar and K\"orner's exponent. 
We have also shown that our proposed algorithm can be used
to calculate cutoff rate and the rate distortion function. 
}
\conclusion
\section*{Acknowledgment}
This work was supported in part by JSPS 
KAKENHI Grant Numbers 23360172, 25820162 and K16000333.



\section*{\empty}
\appendix

\proofprII
\proofPrIII
\prooflmII
\prooflmIII
\proofPropCutoffrate


\end{document}